\documentclass[a4paper,12pt,numbers,sort&compress]{article}

\pdfoutput=1

\usepackage[paper=letterpaper,margin=1in]{geometry}

\usepackage{amsmath,amssymb,amsthm,amsfonts,epsfig,cite,setspace,bigstrut,longtable,array,breqn,url,color}

\usepackage{caption,setspace}

\usepackage{tikz}
\usepackage{mathrsfs}
\usetikzlibrary{arrows}

\usepackage{pdfpages,lipsum}

\begin{document}



\vskip 0.25in

\newcommand{\todo}[1]{{\bf ?????!!!! #1 ?????!!!!}\marginpar{$\Longleftarrow$}}
\newcommand{\sref}[1]{\S~\ref{#1}}
\newcommand{\nn}{\nonumber}
\newcommand{\tr}{\mathop{\rm Tr}}
\newcommand{\comment}[1]{}

\newcommand{\cM}{{\cal M}}
\newcommand{\cW}{{\cal W}}
\newcommand{\cN}{{\cal N}}
\newcommand{\cH}{{\cal H}}
\newcommand{\cK}{{\cal K}}
\newcommand{\cZ}{{\cal Z}}
\newcommand{\cO}{{\cal O}}
\newcommand{\cP}{{\cal P}}
\newcommand{\cR}{{\cal R}}
\newcommand{\cA}{{\cal A}}
\newcommand{\cB}{{\cal B}}
\newcommand{\cC}{{\cal C}}
\newcommand{\cD}{{\cal D}}
\newcommand{\cE}{{\cal E}}
\newcommand{\cF}{{\cal F}}
\newcommand{\cT}{{\cal T}}
\newcommand{\cV}{{\cal V}}
\newcommand{\cX}{{\cal X}}
\newcommand{\IA}{\mathbb{A}}
\newcommand{\IP}{\mathbb{P}}
\newcommand{\IQ}{\mathbb{Q}}
\newcommand{\IH}{\mathbb{H}}
\newcommand{\IK}{\mathbb{K}}
\newcommand{\IR}{\mathbb{R}}
\newcommand{\IC}{\mathbb{C}}
\newcommand{\IF}{\mathbb{F}}
\newcommand{\IV}{\mathbb{V}}
\newcommand{\II}{\mathbb{I}}
\newcommand{\IZ}{\mathbb{Z}}
\newcommand{\re}{{\rm~Re}}
\newcommand{\im}{{\rm~Im}}

\newcommand{\tmat}[1]{{\tiny \left(\begin{matrix} #1 \end{matrix}\right)}}
\newcommand{\mat}[1]{\left(\begin{matrix} #1 \end{matrix}\right)}

\let\oldthebibliography=\thebibliography
\let\endoldthebibliography=\endthebibliography
\renewenvironment{thebibliography}[1]{%
\begin{oldthebibliography}{#1}%
\setlength{\parskip}{0ex}%
\setlength{\itemsep}{0ex}%
}%
{%
\end{oldthebibliography}%
}

\newtheorem{theorem}{\bf THEOREM}
\newtheorem{proposition}{\bf PROPOSITION}
\newtheorem{observation}{\bf OBSERVATION}
\newtheorem{definition}{\bf DEFINITION}
\def\theequation{\thesection.\arabic{equation}}

\newcommand{\setall}{\setcounter{equation}{0}}
\renewcommand{\thefootnote}{\fnsymbol{footnote}}

\begin{titlepage}

~\\
\vskip 1cm
\begin{center}
{\Large {\bf Universes as Big Data}}

\medskip

Yang-Hui He\footnote{
hey@math.ox.ac.uk\\
Invited review for IJMPA, based on various colloquia, seminars and conference talks in the 2019-2020 academic year.
}

\renewcommand{\arraystretch}{0.5} 
{\small
{\it
\begin{tabular}{rl}
  ${}^{1}$ &
  Merton College, University of Oxford, OX14JD, UK\\
  ${}^{2}$ &
    Department of Mathematics, City, University of London, EC1V 0HB, UK\\
  ${}^{3}$ &
    School of Physics, NanKai University, Tianjin, 300071, P.R.~China
\end{tabular}
}}
\renewcommand{\arraystretch}{1.5} 

\end{center}

\vspace{10mm}

\begin{abstract}
We briefly overview how, historically, string theory led theoretical physics first to precise problems in algebraic and differential geometry, and thence to computational geometry in the last decade or so, and now, in the last few years, to data science.
Using the Calabi-Yau landscape - accumulated by the collaboration of physicists, mathematicians and computer scientists over the last 4 decades - as a starting-point and concrete playground, we review some recent progress in machine-learning applied to the sifting through of possible universes from compactification, as well as wider problems in geometrical engineering of quantum field theories.
In parallel, we discuss the programme in machine-learning mathematical structures and address the tantalizing question of how it helps doing mathematics, ranging from mathematical physics, to geometry, to representation theory, to combinatorics, and to number theory.

\end{abstract}

\end{titlepage}

\tableofcontents

\section{Introduction}
The twentieth century has firmly established that the correct language of fundamental theoretical physics is that of algebraic/differential geometry/topology, in all four combinations of these pairs of adjectives and nouns.
Gravity and space-time, should be thought of as the metric and curvature of Riemannian manifolds; elementary particles, irreducible representations of the Lorentz group and gauge connections of appropriate principal Lie-group fibrations, etc.~(q.v.~an attempted modern summary in \cite{Yang:2019kup}).
In some sense, string theory is a brain-child of this tradition.
Whether she will stand as the ultimate theory of everything remains to be seen, but her r\^ole in both bearing the torch and ploughing the field of the conversations between mathematics and physics is unquestionable.

The twenty-first century, with the dramatic progress in computing power and techniques, is bringing a new interlocutor to this dialogue. Already, in the first decade, software such as Macaulay2 \cite{m2}, Singular \cite{singular}, GAP \cite{gap}, MAGMA \cite{magma}, and the umbrella project of SageMath \cite{sage} (launched in 2005), and the increasing online mathematical databases -- \cite{grdb,knots,lmfdb} to name but a few -- are aiding pure mathematical research in an increasingly prominent way (q.v..~launch of ICMS in 2006 \cite{icms}).
In parallel -- and this is of course in tandem with experimental physics whose reliance on and interaction with computers has a rich history of its own -- theoretical physics, and string theory in particular, has benefited from an algorithmic outlook \cite{Kreuzer:2000xy,Gmeiner:2005vz,stats,Anderson:2007nc}.
  
In our present era of Big Data and AI, it is inevitable that machine-learning should have an ever-increasing presence in the second decade \cite{He:2017aed,He:2017set,Krefl:2017yox,Ruehle:2017mzq,Carifio:2017bov}.
The purpose of this talk \cite{talk}, aimed at a general physics audience, is to give an overview of some of this activity in the last few years, especially in the context of machine-learning applied to the string theoretic and geometric landscape, as well as to other mathematical structures within and beyond geometry (cf.~an attempted pedagogical introduction in \cite{He:2018jtw}).
I will take a somewhat historical approach and start with Calabi-Yau data as a concrete playground; this is mainly due to the vastness of the subject in physics and in mathematics, which has consequently led to an abundance of data. The methodology should be applicable to much more general situations, under the rubric of machine-learning mathematical structures \cite{He:2017aed,MLmaths,TC}.

The organization of this talk is as follows.
We begin by reviewing how 2 parallel traditions, one in theoretical physics and one in pure mathematics, converging around 1980s and both leading to the study of complex manifolds and K\"ahler geometry.
This confluence initiated a concerted effort to construct Ricci-flat such spaces, viz., Calabi-Yau manifolds, in the last decade of the 20th century, continuing into the 21st, developing into an explosion of data.
In \S2, we take an overview of this mathematical data, being encouraged by its availability and plenitude, somehow daunted by the complexity of the algorithms needed to process them, and compelled by a thirst for techniques from the ``Big Data'' revolution and AI research.
We then offer the audience an invitation, in \S3, to some modern data science, focusing on machine-learning and how it may be applied to problems in string phenomenology as well as algebraic geometry. 
We conclude with an outlook and report on some recent results in machine-learning fundamental structures in various branches of mathematics and relations to physics.

\section{Trichotomy and Triadophilia}
It is well known that string theory is a unified theory of gravity and elementary particles in high dimensions.
Shortly after the First Revolution in 1984 with anomaly cancellation \cite{Green:1984sg} and the discovery of the heterotic string \cite{Gross:1984dd}, the subject of ``string phenomenology'' was born \cite{Candelas:1985en}.
The reason for this excitement was that all at once, there was an anomaly-free quantum field theory which naturally contained the graviton as well as the $E_8$ gauge group.
In other words, it presented a unified theory of quantum gravity - albeit in 10 space-time dimensions - which also, via the embedding $SU(3) \times SU(2) \times U(1) \subset SU(5) \subset SO(10) \subset E_6 \subset E_8$, could give rise to the standard model. 

\subsection{Complex, K\"ahler, Ricci-flat}
The solution of \cite{Candelas:1985en} was to take inspiration from Kaluza-Klein \cite{KK} and treat the extra $10-4=6$ dimensions as small and space-like, in a {\it compactification} scenario with a 6-manifold $M_6$ on top of each point in our 4-dimensional space-time. Further conditions of supersymmetry \footnote{
Whilst it remains to be seen whether there is supersymmetry in Nature, it is undisputed from a theoretical perspective that quantum field theory with supersymmetry (SUSY) has much richer and tamable structure.
A good (and in some sense rigourous) analogy would be that doing mathematics over $\IR$ is difficult, and this is ameliorated by working over $\IC$, the unique (commutative) algebraic closure.
So too, do the theorems of Coleman-Mandula \cite{coleman} and Haag-\L{}opusza\'{n}ski-Sohnius \cite{HLS}, guarantee SUSY as the unique extension to Poincar\'{e} symmetry in a field theory.
} and vacuum Einstein solutions constrained the 6-manifold to be (1) complex, (2) K\"ahler and (3) Ricci-flat, i.e., respectively 
(1) a complex 3-fold, (2) the metric comes from a scalar potential $g_{\mu \bar{\nu}}(z, \bar{z}) = \partial_\mu \overline{\partial}_{\overline{\nu}} K(z, \overline{z})$ and (3) the Ricci curvature for  $g_{\mu \bar{\nu}}$ vanishes.
We emphasize that this is the simplest solution.
In general, one has to solve the so-called {\it Hull-Strominger} system \cite{Hull-Strominger}, which would lead to a much wider variety of possible $M_6$. As we shall emphasize later, the Calabi-Yau landscape is only a corner of possible compactification scenarios.

To the theoretical physicist in the mid-1980s, perhaps only the word ``Ricci-flat'' was, because of general relativity,  familiar.
Meanwhile, for the mathematical community, this was also rather avant-garde.
The story goes back to classical results of Euler, Gau\ss, and Riemann.
Consider a surface $\Sigma$ - we usually think of a sphere $S^2$ or the surface of a doughnut $T^2$ - and its possible {\bf topological types}, i.e., equivalences up to topology.
Restricting to the cases of smooth, compact (no punctures or boundaries) and orientable (nothing like Klein bottles or M\"obius strips) surfaces, the familiar shapes, $S^2$, $T^2$, and those with increasing number of ``holes'' (genus) are all there is: any smooth, compact, orientable surface can be deformed continuously (topologically homeomorphic) to one of these. 
This single non-negative integer, the genus $g(\Sigma)$, classifies the topology of $\Sigma$.
Closely related is the quantity $\chi(\Sigma) = 2 - 2 g(\Sigma)$, called the {\bf Euler characteristic} or Euler number.
A high-light of the geometry of the 18th-19th centuries is the chain of equalities
\begin{equation}\label{eulerchain}
 2-2g(\Sigma) = \chi(\Sigma) =
\sum\limits_{i = 0}^{\dim_{\IR} \Sigma = 2} (-1)^i b^i (\Sigma) = \frac{1}{2\pi}\int_\Sigma R
\ , 
\end{equation}
where one proceeds, from left to right, 
from topology, to combinatorics, to Gau\ss' Theorema Egregium for differential geometry.
Here, $b^i$ are the {\bf Betti numbers}, counting the number of cycles in dimension $i$ and $R$ is the (Ricci) curvature.
This whole setup above can be complexified where our 2-manifolds -- named {\it Riemann surfaces} -- become complex 1-folds, named {\it complex curves}. 
Furthermore, $\Sigma$ are not just complex, but are also K\"ahler -- one can check that the complex (Hermitian) metric on all such surfaces as complex 1-folds comes from a single potential.

To the equalities \eqref{eulerchain}, early 20th century added another, viz., $\chi(\Sigma) = [c_1(T_{\Sigma})] \cdot [\Sigma]$, where the last integral is re-interpreted as intersection theory between cohomology (here the Chern class $c_1$) and homology (here the class of the manifold). All of these are special cases of the index theorem of Atiyah-Singer and Grothendieck-Riemann-Roch (q.v.~\cite{hartshorne}) which are applicable to spaces of arbitrary dimension.

Having curvature controlling topology also gives us a natural {\bf trichotomy}, which for $\Sigma$, is part of the Riemann Uniformization Theorem. Specifically, in complex dimension 1, we have
\begin{equation}\label{riemann}
R \quad \left\{
\begin{array}{ll}
> 0 :& g=0, \mbox{ Spherical Geometry } \\
= 0 :& g=1, \mbox{ Flat Torus } \\
< 0 :& g>1, \mbox{ Hyperbolic Geometry } \\
\end{array} \right.
\end{equation}
Note that the $R \geq = 0$ cases are finite in topological type and the $R < 0$ cases are infinite.

Much of modern geometry is concerned with generalizing this beautiful story of complex dimension 1 to higher dimensions.
Expectedly, the situation is much more involved and many questions still remain open conjectures.
Nevertheless, for K\"ahler manifolds a conjecture of Calabi \cite{calabi} dating to the 1950s does give the analogue of \eqref{riemann}: essentially, it states that $c_1$, the first Chern class, uniquely controls the Ricci curvature for the K\"ahler metric.
It was not until the Fields-Medal-deserving work in 1978 by Yau \cite{yau} that this was settled.

Fortuitously, Strominger, one of authors of \cite{Candelas:1985en} was visiting Yau at the IAS in 1985 and were neighbours. Thus, the object onto which physicists stumbled -- Ricci-flat, K\"ahler manifolds -- through string compactification, had the world-expert literally next door.
In fact, such spaces were named {\bf Calabi-Yau} manifolds by the physicists.

\subsection{Low-Energy Physics}
Not only did \cite{Candelas:1985en} constrain the compactification manifold, they also established a dictionary between
\[
\mbox{``Geometry of $X_6$ $\longleftrightarrow$ physics of $\IR^{1,3}$.''}
\]
Purely working from the group theory, the tangent bundle with its $SU(3)$ structure breaks the $E_8$ to an $E_6$ (SUSY) GUT theory.
We will skip the details, but basically for the fundamental fermions (the choice of which is anti-generation and generation is by convention):
\begin{equation}\label{gens}
\begin{array}{rl}
\mbox{generations of particles} & \sim h^{2,1}(X) \ , \\
\mbox{anti-generations of particles} & \sim h^{1,1}(X)  \ ,
\end{array}
\end{equation}
where $h^{p,q}$ are the {\bf Hodge numbers} of the Calabi-Yau manifold $M_6$, the complexified version (hence the double index, for complex and conjugate) of the Betti numbers mentioned earlier.
The alternating sum of Betti numbers to the Euler number generalizes to a double sum, and in particular $\chi(M_6) =  2(h^{1,1}(M_6) - h^{2,1}(M_6))$.
Since there are 3 generations of fermions, one of the original constraints of \cite{Candelas:1985en} is that
\begin{equation}\label{TX3}
\left| h^{2,1}(X) - h^{1,1}(X) \right| = 3 \Rightarrow \chi(X) = \pm 6 \  .
\end{equation}
Finding compact, smooth Calabi-Yau 3-folds with Euler number $\pm 6$ was perhaps historically the first concrete challenge physicists gave to the algebraic geometry community.

\paragraph{Disclaimer: }
It must be emphasized that \eqref{gens} is only for the so-called standard embedding for the heterotic string to get to $E_6$-GUT theories, the field has since evolved to far beyond merely computing Hodge numbers, but to computing equivariant cohomology of stable bundles (cf.~\cite{Braun:2005nv,Bouchard:2005ag,Anderson:2012yf,Constantin:2018xkj}).
In addition, there is a myriad of phenomenological approaches from other string/M-/F-theoretic constructions, which constitute the vastness of the ``string landscape'', the review of which is not our present intent.
The reader is referred to the wonderful textbooks \cite{books} in general, and to the classic \cite{Candelas:1987is} for an introduction to complex geometry for physicists, and, in the context of Calabi-Yau spaces, to \cite{He:2020bfv} for a brief invitation and \cite{He:2018jtw} for a pedagogical textbook.

Although the physics community is no longer searching for manifolds with property \eqref{TX3}, there is an entire programme, especially led by Candelas, to look for Calabi-Yau manifolds of {\it small} Hodge numbers \cite{Candelas:2016fdy} which have interesting mathematics of its own.
In any event, the search for a geometric interpretation, or origin, of 3 generations of particles has been dubbed ``Triadophilia'' \cite{Candelas:2007ac}.
In a way, the dictionary started by \eqref{gens}, where properties of our universe are purely phrased in the geometry of some manifold, is an elegant modern realization of Kepler's famous adage: ``Ubi materia, ibi geometria'' \footnote{
``Where there is matter, there is geometry,'' from Johannes Kepler's Thesis XX from {\it De fundamentis astrologiae certioribus} 1602.
}.
Perhaps for this reason by itself, it is worth studying string theory as a theory of physics, let alone its cross-fertilizations to mathematics and - as we will see - data science.

\subsection{Early Constructions}
One of the first questions which the physicists asked Yau was, indubitably, ``how to construct an explicit Calabi-Yau 3-fold?''
It is curious that students in theoretical physics are taught differential geometry first, before algebraic geometry, whereas the fundamental ideas of the latter - vanishing loci of polynomials - are certainly more familiar than that of the former - local patches and differentiable transition functions.
We know how to construct shapes from Cartesian geometry since our early school days.
For example, a quadratic equation in two real variables $(x,y)$ is a conic section, such as a circle.
Thus, the vanishing locus on a quadratic polynomial in two real variables gives a $2-1=1$ dimensional real manifold in an ambient $\IR^2$.
We have just created a simple {\bf algebraic variety}.

Now, we are looking for complex manifolds, we thus construct them as the zero-locus of multiple polynomials in multiple complex variables.
In this way, a Calabi-Yau 1-fold,  a Riemann surface of zero curvature, viz.~the torus $T^2 = S^1 \times S^1$,
is realized as a cubic in two complex variables given by  the so-called Weierstra\ss\ equation
$T^2 \simeq \{x,y \in \IC | y^2 = x^3 - g_2 x - g_4\} \subset \IC^2$,
where $g_{2,4}$ are complex constants.
One can check by writing out $(x,y)$ in their real and imaginary parts, and the Weierstra\ss\ equation becomes 2 real constraints in 4 real variables, which we can numerically plot by Monte Carlo to see a torus emerge.
Next, compactness can be ensured by including the point at infinity, where $(x,y) = (\infty, \infty)$.
One can do this by so-called {\it projectivization} where instead of $\IC^2$,  we introduce one more complex coordinate, $z$ such that any point $(x,y,z) \in \IC^3$ is identified with the scaled $\lambda (x,y,z)$ for non-zero $\lambda \in \IC$.
This scale-invariance brings the point at infinity to a finite point, rendering the resulting ambient space and the subsequent torus compact.

What we have done is to construct, from $\IC^3$ with coordinates $(x,y,z)$, the complex projective space $\IC\IP^2$ with {\bf homogeneous} coordinates $[x:y:z]$.
More formally, we define $\IC\IP^n$ from $\IC^{n+1}$ with coordinates $(z_0, z_1, \ldots, z_n)$ as the quotient by the equivalence relation $\sim$
\begin{equation}\label{cpn}
 \IC\IP^n := \IC^{n+1} \backslash \{ \vec{0} \} \bigg/  (z_0,z_1,\ldots,z_n) \sim  \lambda  (z_0,\ldots,z_n)  \ , \qquad
  \lambda \in \IC \backslash \{0\} \ .
\end{equation}
The complex $n$-fold $\IC\IP^n$ is smooth, with $n+1$ homogeneous coordinates.

Thus, the Calabi-Yau 1-fold is realized as a so-called projective algebraic variety inside $\IC\IP^2$
\begin{equation}\label{weierstrass}
\{[x:y:z] |- y^2z + x^3 - 4g_2 xz^2 - g_4z^3 = 0 \} \subset \IC\IP^2 \ ,
\end{equation} 
a homogeneous cubic in the homogeneous coordinates $[x:y:z]$ of $\IC\IP^2$.
Luckily, complex projective space and the zero loci of any number homogeneous polynomials therein, are guaranteed to be K\"ahler. For $\IC\IP^n$, the K\"ahler metric explicitly comes from the famous Fubini-Study potential $\log( 1+ \sum_i |z_i|^2)$.
This construction is valid in general: the hypersurface defined by a homogeneous polynomial of degree $n+1$ in $\IC\IP^n$ is a Calabi-Yau $(n-1)$-fold.
Thus we arrive at our first, and perhaps most famous, example of a Calabi-Yau 3-fold: the quintic hypersurface in $\IC\IP^4$.
There are many degree 5 monomials  one could compose of 5 coordinates, the most well-studied is the so-called Fermat quintic:
\begin{equation}
Q := \{ x_0^5 + x_1^5 + x_2^5 + x_3^5 + x_4^5 = 0 \} \subset
\IC\IP^4_{[x_0:x_1:x_2:x_3:x_4]} \ .
\end{equation}
What are the topological numbers of $Q$? The Hodge numbers turn out to be
$h^{2,1}(Q) = 101$ and $h^{1,1}(Q) = 1$ so that $\chi(Q) = 2(1-101) = -200$.

Immediately, we also obtain 4 close relatives.
Consider the intersection of 2 cubics in $\IC\IP^5$; this is a complete intersection in that the number of defining polynomials - here 2 - is equal to the codimension -- i.e., the dimension of the ambient $\IC\IP^5$ minus the dimension of the required manifold, $5 - 3 = 2$.
We denote this as $[5 | 3 , 3]$, much as we could denote the quintic as $[4 | 5]$.
Note that the number of to the left of the bar is 1 less than the row-sum (Calabi-Yau condition) and also 3 more than the number of columns (complete intersection condition).
A simple integer partition shows that there are 5 possibilities in total, including the quintic, viz.,
\begin{equation}\label{cyclics}
[4 |5 ], \ [5 | 3,3], \ [5 | 2,4], \ [6 | 2,2,3], \ [7 | 2,2,2,2] \ .
\end{equation}
These are called {\em cyclic} Calabi-Yau 3-folds, and are the only ones as complete intersections in a single projective space.

We need to emphasize that complete intersections are rare and most algebraic varieties are {\it not} so.
In fact, there is a general result that (see \cite{GHJ})
\begin{theorem}
All K\"ahler 3-folds can be realized as vanishing loci of systems of polynomials in $\IC\IP^7$. 
\end{theorem}
Therefore, one could in principle write all sorts of (non-complete-intersection) polynomials in 8 homogeneous variables and sift out the Calabi-Yau ones; but this is highly impractical.

Of the 5 immediate ones, none has the property \eqref{TX3}, so the community turned to more general constructions.
Again, as mentioned earlier, today physicists are no longer limited to \eqref{TX3} and \eqref{cyclics} have all been met with renew zest. In the late 1980s, however, a different path was undertaken, and an industry of subsequent generalizations to \eqref{cyclics} was initiated:
\begin{description}

\item[CICYs]
The first generalization is to take, instead of a single $\IC\IP^n$, a product $A$ of projective spaces.
That is, let $A = \IC\IP^{n_1} \times \ldots \times \IC\IP^{n_m}$, of dimension $n = n_1 +n_2 + \ldots + n_m$ and each having homogeneous coordinates $[x_1^{(r)}:x_2^{(r)}:\ldots:x_{n_r}^{(r)}]$ with the superscript $(r) = n_1, n_2, \ldots, n_m$ indexing the projective space factors.
The Calabi-Yau 3-fold is then defined as  the complete intersection of $K = n-3$ homogeneous polynomials in the coordinates $x_j^{(r)}$.
Succinctly \footnote{
Importantly, the Chern classes and the Euler number can be read off the matrix configuration explicitly.
\comment{
We have that $c_1(T_X) = 0$ and moreover,
\begin{equation}\label{chernCICY}
c_2^{rs}(T_X) = \frac12 \left[ -\delta^{rs}(n_r + 1) + 
  \sum_{j=1}^K q^r_j q^s_j \right] \ , \quad
c_3^{rst}(T_X) = \frac13 \left[\delta^{rst}(n_r + 1) - 
  \sum_{j=1}^K q^r_j q^s_j q^t_j \right] \ , 
\end{equation}
where we have written the coefficients of the total Chern class
$c = c_1^r J_r + c_2^{rs} J_r J_s + c_3^{rst} J_r J_s J_t$ explicitly, with $J_r$ being the K\"ahler form in $\IP^{n_r}$.
The triple-intersection form $d_{rst} = \int_X J_r \wedge J_s \wedge J_t$ is a totally symmetric tensor on $X$ and the Euler number is simply $\chi(X) = d_{rst} c_3^{rst}$.
}
The individual terms $(h^{1,1},h^{2,1})$, however, cannot be deduced from the configuration matrix directly.
This is one of the short-comings of the index theorem: the integral of curvature and the intersection of the Chern classes give only the alternating sum (Euler number) in (co-)homology, but not the individual terms.
},  this information
can be written into an $m \times K$ configuration matrix which generalize \eqref{cyclics}:
\begin{equation}\label{cicy}
X = 
\left[\begin{array}{c|cccc}
  \IC\IP^{n_1} & q_{1}^{1} & q_{2}^{1} & \ldots & q_{K}^{1} \\
  \IC\IP^{n_2} & q_{1}^{2} & q_{2}^{2} & \ldots & q_{K}^{2} \\
  \vdots & \vdots & \vdots & \ddots & \vdots \\
  \IC\IP^{n_m} & q_{1}^{m} & q_{2}^{m} & \ldots & q_{K}^{m} \\
  \end{array}\right]_{m \times K \ ,}
\quad
\begin{array}{l}
K = \sum\limits_{r=1}^m n_r-3 \ , \\
\sum\limits_{j=1}^K q^{r}_{j} = n_r + 1 \ , \ \forall \; r=1, \ldots, m \ .
\end{array}
\end{equation}

These manifolds defined by \eqref{cicy} were called CICYs (complete intersection Calabi-Yau manifolds) and were explicitly constructed by Candelas et al.~\cite{cicy} (q.v.~H\"ubsch's classic book \cite{hubschbook}) in the early 1990s.
The combinatorial problem for these integer matrices turned out to be rather non-trivial and one of the most powerful super-computers then available, the one at CERN,  was recruited.
To our knowledge, this might have been the first ``data-base'' in algebraic geometry.

Up to trivial equivalence such as row/column permutations as well as non-trivial ones such as so-called splitting, CICYs were shown to be finite in number, a total of 7890 configurations, with a maximum of 12 rows, a maximum of 15 columns, and all having entries $q_j^r \in [0,5]$.
There are 266 distinct Hodge pairs $(h^{1,1},h^{2,1}) = (1,65), \ldots, (19,19)$,  giving 70 distinct Euler numbers $\chi \in [-200,0]$.

\item[WP4s]
Noticing that the CICY data is rather skewed in that all Euler numbers were non-positive, the constructions went on.
The reason for this is that physicists knew about mirror symmetry by then, one of whose most salient features is the exchange  of $h^{1,1} \leftrightarrow h^{2,1}$, which would flip the sign of $\chi$.
Another way to generalize projective space is to introduce weights \footnote{
In fact, products of projective spaces can also be thought of as a weighted projective space with vector-valued grading.
}.
That is, one takes {\it weighted} projective space $\IC\IP^4_{[d_0:\ldots:d_4]}$ as the ambient space $A$, which generalizes \eqref{cpn} by having integer ``weights'' $(d_0,d_1,d_2,d_3, d_4) \in \IZ_{+}$ as
\begin{equation}\label{wp4}
\IC\IP^4_{[d_0:\ldots:d_4]} := \IC^5 \backslash \{ \vec{0} \} \bigg/ \left( (z_0,z_1,\ldots,z_4) \sim (\lambda^{d_0} z_0,\ldots,\lambda^{d_4} z_4) \right)\ , \qquad
\lambda \in \IC \backslash \{0\}\ . 
\end{equation}
Taking all weights $d_i=1$ is the ordinary $\IC\IP^4$.
As with $Q$, if we embed a hypersurface of degree $d_0 + d_1 + \ldots + d_4$ into $\IC\IP^4_{[d_0:\ldots:d_4]}$, it defines a CY$_3$.
The classification of such manifolds was performed in \cite{Candelas:1989hd} and a total of 7555 is found,
with 2780 distinct Hodge pairs and a more balanced $\chi \in [-960, 960]$.

\item[Reflexive Polytopes]
The next systematic generalization of weighted projective space is a {\bf toric variety}, which, instead of having a single list of weights as in \eqref{wp4}, has a list of $m$ weights (giving a so-called charge-matrix) acting on $\IC^{n+m}$ to give an $n$-fold.
Based on the theorem of Batyrev-Borisov \cite{BB}, Kreuzer and Skarke spent almost a decade
explicitly constructing such Calabi-Yau manifolds,
culminating in the early 2000s with the construction of the most extensive database of CY$_3$ so far, the {\bf Toric Hypersurfaces} \cite{Kreuzer:2000xy}.

In brief, the ambient space is a toric 4-fold $A$, constructed from an integer polytope $\Delta \subset \IR^4$ which is {\bf reflexive}, meaning that $\Delta$ has a single interior point (which can be taken to be the origin) and all bounding hyperplanes are distance 1 from this point.
Furthermore, a particular hypersurface in the toric variety $A$ is a CY with defining equation of the CY$_3$ is given by
\begin{equation}\label{bbHypersurface}
X = \{
\sum\limits_{\vec{m} \in \Delta} c_{\vec{m}} \prod\limits_{j=1}^k x_j^{\vec{m} \cdot \vec{v}_j + 1} = 0
\} \subset A \ ,
\end{equation}
with $x_j$ coordinates of the ambient toric 4-fold, $c_{\vec{m}}$ complex coefficients, and $\vec{v}_j$ the (integer) vertices of $\Delta^\circ$.
The weighted $\IC\IP^4$ hypersurfaces are special cases of \eqref{bbHypersurface}.

Thus the question of finding toric hypersurface CY$_3$ is the classification of reflexive integer 4-polytopes (up to  $SL(4; \IZ)$, under which the toric 4-folds are equivalent).
In $\IR^{1}$, there is trivially 1 reflexive polytope (the pair of points $\pm 1$).
In $\IR^2$, it is known at least to 19-th century mathematics, that there are 16 reflexive polygons up to $SL(2; \IZ)$.
Unfortunately (and perhaps shockingly), the next number is already unknown until the work of Kreuzer-Skarke.
They found 4319 reflexive polyhedra in $\IR^3$.
For $\IR^4$, 6 months of computation on the best computer available to the late 1990s gave an astounding 473,800,776.
Each of these gives \footnote{
It should be emphasized that most of these toric ambient spaces $A$ (as with weighted $\IC\IP^4$) are {\it not smooth}, and requires smoothing or resolution of singularities: different resolutions give rise to potentially different CY$_3$s. 
Thus, the actual number of Calabi-Yau 3-folds from this construction is estimated to be many orders of magnitude larger.
For a given $\Delta$, the Hodge pair will be the same, different resolutions will give different intersection numbers and Chern classes.
Up to $h^{1,1} = 7$, this was done exhaustively in \cite{Altman:2014bfa}, while for the highest $h^{1,1} \sim 490$, this was done in \cite{Braun:2017nhi}. The full list of CY$_3$s, after all the resolutions, has been recently estimated to be as large as $10^{10^5}$ \cite{Altman:2018zlc}.
} a hypersurface Calabi-Yau 3-fold.
Thus, our zoo of manifolds increased from 5, to some 10 thousand, and to some half-billion.
Interestingly, the next number, that of reflexive polytopes in $\IR^5$ up to $SL(5;\IZ)$, is unknown.
It would be great to have a generating function for the sequence $1; 16; 4319; 473,800,776; \ldots$.

The KS dataset  produced 30,108 distinct Hodge pairs and $\chi \in [-960, 960]$, with
the extremal values of $\pm 960$ being the weighted $\IC\IP^4$ cases.
No CY construction so far has ever produced an Euler number whose magnitude exceeds 960.
A conjecture of Yau states that the topological type of (connected, smooth, compact) Calabi-Yau manifolds is {\it finite} in every dimension (we already see this in complex dimensions 1 and 2) and it could well be that 960 is the upper bound in dimension 3.
There has been nice parallel directions of work in infinite families of Calabi-Yaus \cite{GWinf,gasparim} beyond topological type such as Gromow-Witten invariants, as well as in zooming in on special corners of small Hodge numbers \cite{Candelas:2007ac,Candelas:2016fdy,HalMikeSmall}.

\end{description}

Thus, by the turn of the century, there is an data-base of Calabi-Yau manifolds whose size is ``big'' even by today's standards.
There is an internet meme, that ``technically, Moses was the first person to download data from the cloud using a tablet'' \cite{moses}.
This amusing anachronism is a fitting analogy to how the age of ``big data'' in theoretical physics and algebraic geometry really goes back to the 1980s.

\section{Data Explosion}
Meanwhile, by the mid to late 1990s, in parallel to the heterotic programme outlined above, the discovery of D-branes \cite{Polchinski:1995mt}, M-theory and $G_2$-compactification \cite{Horava:1996ma,Atiyah:2001qf}, F-theory \cite{Vafa:1996xn,Freview}, AdS/CFT \cite{Maldacena:1997re}, etc., as well as the wealth of dualities linking them begat the Second String Revolution.
As with the First, this gave rise, and is still continuing to engender, a plethora of mathematical data, leading to various estimates of the ``string landscape'' \cite{Kachru:2003aw,stats}, which was already anticipated in \cite{landscape1}.
Numbers such as $10^{500}$ and, as aforementioned, today's $10^{10^{5}}$ began to enter the string and popular psyche.

In some sense, string theory has traded one difficult problem -- the quantization of gravity -- with another: the selection of the right vacuum.
The latter is perhaps of more and certainly increasing interest to pure mathematicians, because the largess of data provides an inspiring playground for generating, testing and proving new conjectures.

Ultimately, whichever scenario one prefers to geometrically engineer (to use the phrase of \cite{Katz:1996fh}) one's preferred quantum field theory, including the standard model, the procedure can be algorithmized.
Indeed, any problem in algebraic geometry (over $\IC$) reduces to finding an appropriate Gr\"obner basis and then to finding (co-)kernels of integer matrices (in a corresponding monomial basis) \cite{m2}.

Take, as an example, AdS/CFT from the point of view of computational geometry: this is a correspondence between a SUSY conformal QFT and a (non-compact) Calabi-Yau cone $M$ over a Sasaki-Einstein manifold $X$.
Moreover specifically, this is a mapping between a quiver representation and the geometric data of $X$ (see e.g., \cite{He:2016fnb} for a quick review).
When $M$ is toric, for instance, the graph data of the quiver and the combinatorial data of $M$ are both amenable to an algorithmic treatment.

\subsection{The Good, The Bad, and The ?}
Taking stock of the progress up to the second decade of this century, we hope to have given the reader a glimpse of how computational and algorithmic geometry has enriched the classical dialogue between physics and mathematics.
The ever increasing number of (freely available) mathematical  databases online (typically of size $\sim 1 - 10$ Gb) is augmented by ever-more efficient software developed to address them (especially the umbrella project of SageMath \cite{sage}) as well as  by the growing power of the personal laptop. This, certainly can be considered ``the Good.''

Unfortunately, most algorithms needed to compute anything, whether it be finding Gr\"obner bases, obtaining triangulations of polytopes, or extracting dual cones, are exponential in complexity.
Thus, if one aims to sift through vacua to find the standard model or to understand the minimal model approach to algebraic varieties, case-by-case checks is impossible, even with the best HPC available.
This, certainly needs to be rendered as ``the Bad''.

While the statistics of the vacuum degeneracy in string theory had been considered in the last decade \cite{stats}, it is only expedient, given the breath-taking speed with which the Big Data Revolution is taking over every aspect of civilization, especially in this decade of the new millenium, that one should apply the most recent techniques to address the landscape of mathematical data.

In many ways, the search of the standard model within the string landscape reminds us of the hunt for exo-planets.
The latter scans the heavens for habitable earths and the former, for universes akin to ours.
The latter accumulates more and more real data with the betterment of technology and the former, theoretical data with furtherance of methodology.
Whether one believes our universe is ``special'' by some anthropic argument, or by a selection principle, or is a mere point in the multiverse, is currently still a matter of debate, but the big data of mathematical universes beckon exploration.

\section{Deep-Learning the Landscape}
A question which instinctively occurred whilst contemplating the big data of universes \cite{He:2017aed,He:2017set}, was that the typical problem in string theory, or, in algebraic geometry for that matter, is of the form
\[
	\stackrel{INPUT}{\fbox{\mbox{integer tensor}}} \longrightarrow \stackrel{OUTPUT}{\fbox{\mbox{integer}}} \ .
\]
That one has integer output, especially in string theory, is because presently much of the field is still at the stage of finding quantities such as number generations, or the charges of particles.
The fact that there is no known non-trivial (compact) Calabi-Yau metric analytically (Yau's proof of the Calabi Conjecture is famously non-constructive and relies on subtle existence statements of Monge-Amp\`ere PDEs) hinders questions such as finding masses \footnote{
We will later address some of the recent advances in numerical metrics and connections.
}.
In geometry, much of the field is concerned with finding topological invariants such as indices or Betti numbers (as mentioned in introduction in Gau\ss' theorema egregium) because when complicated integrals become integers there is usually some deep mathematics going on.
On the other hand, one has integer tensor input is seen in a multitude of examples above: whether we are dealing with polytopes or CICY configurations or quiver adjacency matrices.
Of course, in general cases (such as numerical metrics), the integer condition case be relaxed to numerical tensor input going to some numerical output.

As discussed repeatedly, the machinery of computational geometry has developed sophisticated algorithms to obtain the output from the given input, even though the generic such algorithm is expensive.
Nevertheless, physicists and mathematicians  have bitten the bullet over the last 20 years or so and computed extensive examples. For instance, all Hodge numbers for the 1/2-billion Kreuzer-Skarke Calabi-Yau manifolds have been calculated using combinatorics of polytopes \cite{palp}, likewise, all those for CICYs have been obtained by chasing exact sequences \cite{hubschbook}.

The situation is rather reminiscent of hand-writing recognition, the archetypal problem in machine-learning.
For example, I write 0 to 9 as follows
\begin{equation}
\includegraphics[trim=0mm 0mm 0mm 0mm, clip, width=2in]{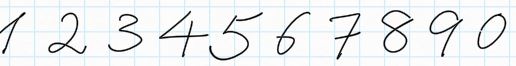}
\end{equation}
and we wish to let the computer recognize them.
The input is an image, which is an $m \times n$ matrix (indexing the pixels in a 2-dimensional grid) each entry of which is a 3-vector of a real value between 0 and 1, denoting the percentage of RGB values. If we only wish to keep gray-scale information, each entry is then a real number between 0 and 1. Or, if we only want black-white, the input is just a binary matrix.
The output is an integer from 0 to 9, called a {\it 10-channel output}.

As mathematicians or theoretical physicists, we might solve this problem by exploiting the geometry and find, say, a clever Morse function as we scan the input matrix row-wise and column-wise and detect the critical points.
This is, of course, very expensive.
What Google or your smart-phone does, is to turn to {\it labeled data}.
Such data has been painfully collected over the years by NIST (National institute of Standards), and look like the following (each is given as, for example, by a $28 \times 28$ pixelated image):
\begin{equation}\label{sampledigits}
\includegraphics[trim=0mm 0mm 0mm 0mm, clip, width=2.5in]{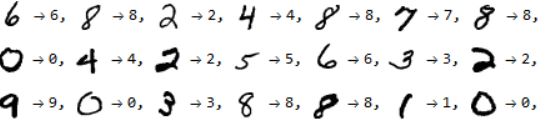} \ldots
\quad
\fbox{\includegraphics[trim=0mm 0mm 0mm 0mm, clip, width=0.5in]{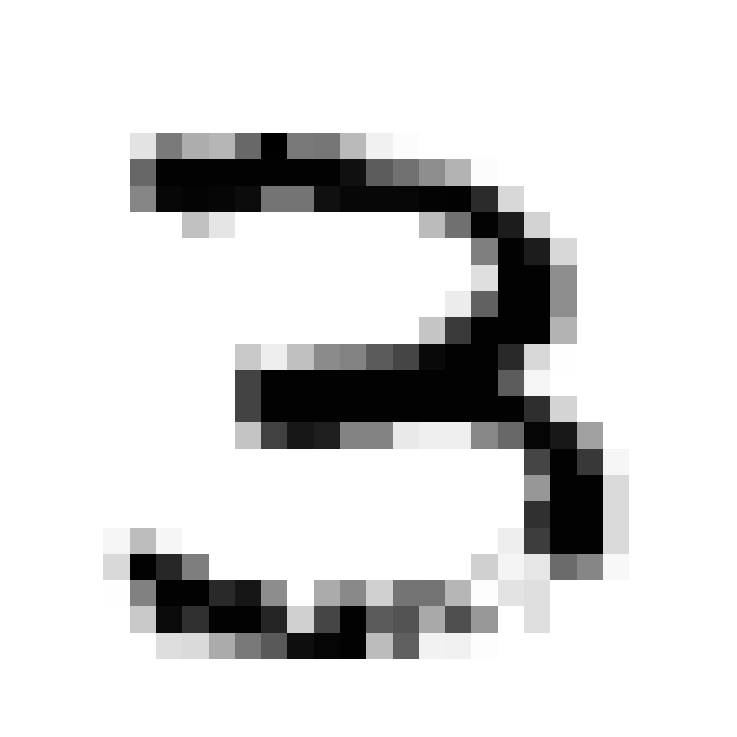}} \quad 
28 \times 28 \times (RGB)
\end{equation}
The difficult part of labeling each image with the correct channel has been done, and still adjusting with new usage by new users.

In summary, what happens is the following:
\begin{description}
\item[Data Acquisition: ] the collection of known cases (input $\to$ output), such as \eqref{sampledigits}, gives us {\it training data};
\item[Machine-Learning: ] setting up some algorithm to optimize parameters which does the classification best;
\item[Data Validation: ] once the machine has ``learnt'' the training data, we can take a set of {\it validation data}, which, importantly, the machine has {\em not} seen before.  This is in the same format as the training data, with given input and output and we check the actual with the predicted outputs.
\end{description}

How different, really, is a problem in algebraic geometry?
For example, computing the Hodge number of a given CICY, after all the work in long exact sequences in cohomology, gives the association rule
\begin{equation}\label{cicy2000}
X={\tiny\left( \arraycolsep=1.4pt\def\arraystretch{1}
\begin{array}{cccccccc}
 1 & 1 & 0 & 0 & 0 & 0 & 0 & 0 \\
 1 & 0 & 1 & 0 & 0 & 0 & 0 & 0 \\
 0 & 0 & 0 & 1 & 0 & 1 & 0 & 0 \\
 0 & 0 & 0 & 0 & 1 & 0 & 1 & 0 \\
 0 & 0 & 0 & 0 & 0 & 0 & 2 & 0 \\
 0 & 1 & 1 & 0 & 0 & 0 & 0 & 1 \\
 1 & 0 & 0 & 0 & 0 & 1 & 1 & 0 \\
 0 & 0 & 0 & 1 & 1 & 0 & 0 & 1 \\
\end{array}
\right)} \ , \quad
h^{2,1}(X) = 22;
\quad
\begin{tabular}{c}\includegraphics[trim=70mm 50mm 100mm 50mm, clip, width=2in]{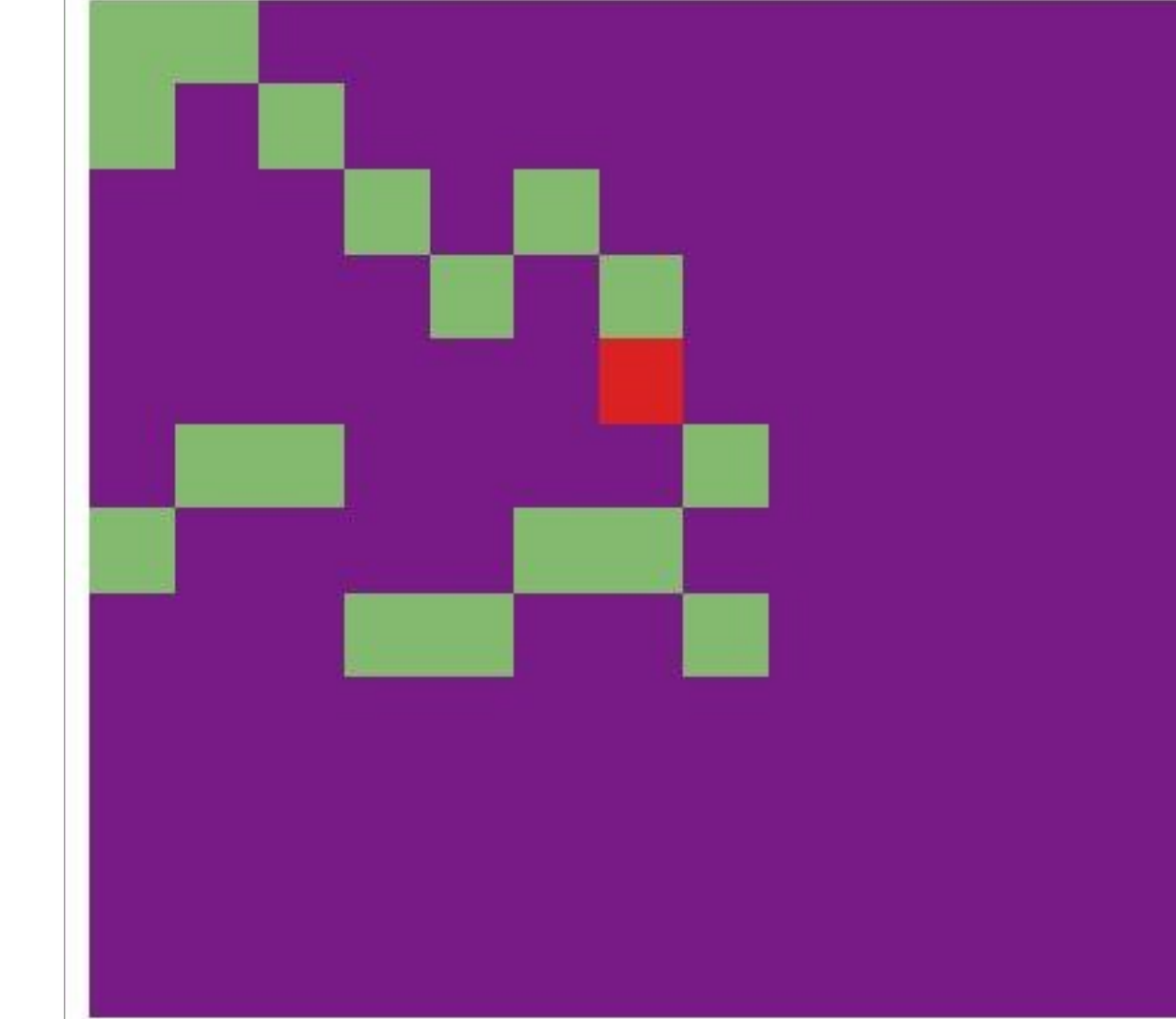}
\end{tabular}
\rightarrow
22 \ .
\end{equation}
To the right, we have purposefully represented the CICY configuration as a pixelated image, since all CICYs
can be embedded, after right-bottom zero-padding, into a $12 \times 15$ integer matrix with entries $\in [ 0, 5]$.
We have 7890 labeled data-points, from which we can take, say, 80\% for training, to be validated on the remaining 20\%.
The programme of machine-learning algebraic geometry was thus initiated \cite{He:2017aed,He:2017set}.

It is timely, that in 2017 (the same year that Sophia, the AI robot, became the first non-human citizen of a country), 4 independent groups were thinking about various aspects of machine-learning the string landscape \cite{He:2017aed,Krefl:2017yox,Ruehle:2017mzq,Carifio:2017bov}.
Thinking back, let us see the sequence of the starting year of annual series of conferences in the string community:
``Strings'' (1986-), `StringPheno'' (2002-), ``NSF String Vacuum Project'' (2006 - 2010), ``String-Math'' (2011-), 
session of stringy mathematics and physics at ``SIAM'' (2014-), and now, ``String-Data'' (2017-).

\subsection{An Invitation to Machine-Learning}
We refer the reader to the now classic introduction to machine-learning in \cite{GBA} as well as a wonderful new monograph for physicists in \cite{Ruehle:2020jrk}.
Here, it is expedient to give a rapid taster of this vast subject.

Contrary to expectations, the field of machine-learning and neural networks goes as far back as cybernetics in the 1940s.
In 1957, the first {\it perceptron} was set up by MIT-Cornell, where a wall of CdS photo-receptors was set up to emulate neurons firing.
In the 1980 - 90s, artificial neural networks went under the philosophy of connectivism where computational power emerged from inter-connectivity.
Slowly the word ``artificial'' disappeared and such algorithms were simply called {\it neural networks} (NNs).
By 2006, the phrase ``Deep'' NN came into being, a term which we will explain shortly.

In general, sorting data into discrete categories is done by {\bf classifiers} and predicting continuous values, {\bf regressors}.
Given data, machine-learning (ML) roughly fall under the headings of {\bf unsupervised}, where patterns are to be extracted, and {\bf supervised} where {\it labeled data}, such as the ones in \eqref{sampledigits} and \eqref{cicy2000}, where the ML algorithms are trained to associate input to output.
Examples of unsupervised ML include clustering analysis, auto-encoders, principle component analyses (PCA), etc., and those of supervised ML include support vector machines (SVM), neural regressors and neural classifiers, etc.
In this talk, I will concentrate, because of the nature of the problem, on supervised ML.

Let us start with a single neuron (the perceptron), which consists of a (ususally analytic) function $f(z_i)$ called the {\it activation function}, for some input tensor $z_i$ with multi-index $i$.
We then consider $f( w_i z_i + b)$ with  weights  $w_i$ and bias $b$.
Typical activation functions include:
(1) Logistic Sigmoid: $\left( 1 +  e^{-x} \right)^{-1}$;
(2) Hyperbolic tangent: $\tanh(x) = \frac{e^{x} + e^{-x}}{e^{x} - e^{-x}}$;
(3) Softplus:  $\log \left( 1 + e^x\right)$, a ``softened'' version of ReLu (Rectified Linear Unit): $\max(0,x)$;
(4) Softmax:  $x_i \rightarrow \frac{e^{x_i}}{\sum_i e^{x_i}}$;
(5) Identity: $x_i \to x_i$ (which, with weights and biases, becomes the general affine transformation).

Given {Training data}: $\cD = \{ (x_i^{(j)}, d^{(j)} \}$ with input $x_i$ and {known output} $d^{(j)}$, we minimize some appropriate {\bf cost/loss function} to find optimal $w_i$ and $b$ (this is the ``learning''). 
Then, with parameters fixed, we can check against {Validation Data}.
Common cost functions include SEL (squared-error-loss)
  \begin{equation}
  SEL := 
  \sum\limits_j \left[ f \left(\sum_i w_i x_i^{(j)} + b\right) - d^{(j)} \right]^2
  \end{equation}
for continuous output, and XC (cross-entropy)
  \begin{equation}
 XE :=
  -\frac{1}{n} \sum\limits_j \left[  d^{(j)} \log f(x^{(j)}) + (1 - d^{(j)}) \log ( 1 - f(x^{(j)}) ) \right]
 \end{equation}
 for discrete (categoritcal) data.
 
The astute reader would recognize that we have done is precisely (non-linear) regression.
With a single neuron, supervised ML is exactly that.
When we link up a multitude of neurons into a directed graph, complexity emerges through connectivity in a gestalt-philosophical way; this is the NN.
A common type of NN is when the graph organizes into ``layers'' as in
\begin{equation}
\begin{array}{cc}
\rotatebox[origin=c]{90}{$\longleftarrow$ width $\longrightarrow$}
 & \begin{array}{c}\includegraphics[trim=0mm 0mm 0mm 0mm, clip, width=4.5in]{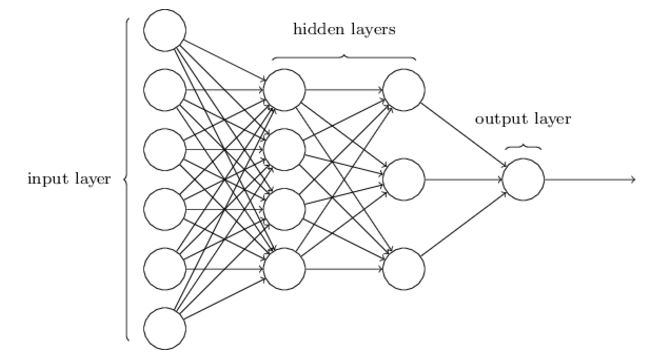}\end{array} \\
 & \longleftarrow \mbox{depth} \longrightarrow
 \end{array}
\end{equation}
which is called a {\it forward-feeding} NN (or, a more dated acronym, MLP, for multi-layer perceptron).
The MLP is composed of an input layer, an output layer, and a number of hidden layers.
The total number of layers is called the {\bf depth} and the rough number of neurons per layer, the {\bf width}.
ML with large depth NNs is, for obvious reasons, called {\it deep learning}.

The precise choice of activation functions and inter-connectivity of the NN is called the {\it architecture}.
The various parameters - not the variables like weights and biases to be optimized during training - such as depth, width, learning-rate (this the step-size for any gradient descent method using for finding minima), batch-size (the training data is usually passed in batches at a time), etc., are called {\it hyper-parameters}.
As one can imagine, there is a variety of {\bf universal approximation theorems} which essentially state that for sufficiently large width, or depth, any output can be approximated to arbitrary precision. In fact, a forward-feeding fully-connected NN with only ReLU activation is good enough to approximate any integrable function.

As with all models of statistical prediction, it is good to have a measure of ``goodness of fit''.
Some standard ones are as follows
\begin{description}
\item[Na\"{\i}ve Precision: ]
This is particularly useful when the output is discrete (and belonging to a relatively small number of categories), and we simply compute the percentage of agreed cases between the predicted and actual.

\item[R-squared: ]
For continuous output, suppose on validation dataset $\cV = \{ x_i^{(j)} \longrightarrow { d^{(j)} } \}_{j = 1,2,\ldots, m}$, the predicted values are $\{ x_i^{(j)} \longrightarrow { \hat{d}^{(j)} } \}_{j}$.
Then the Coefficient of Determination, or simply R-squared, is defined to be
$R^2 := 1 - \frac{SS_{\mbox{res}}}{SS_{\mbox{tot}}}$, where data variance is $SS_{\mbox{tot}} := \sum_j (d^{(j)} - \overline{d^{(j)}})^2$ for mean $\overline{d^{(j)}}$, and residual sum of squares is
$SS_{\mbox{res}} := \sum_j (d^{(j)} - \hat{d}^{(j)})^2$.
A bad fit is when $R^2$ is close to 0, and a perfect fit, when $R^2=1$.

\item[Confusion Matrix: ]
For discrete output (say $n$ categories), we can establish an $n \times n$ matrix with the $(i,j)$-th entry being the number of cases predicted to be $j$ while the actual value is $i$.
Ideally, we wish this to be a diagonal matrix.
A measure of how close to the diagonal is the {\bf Matthews' $\phi$-coefficient} defined to be $\sqrt{\chi^2/n}$ where $\chi^2$ is the Chi-square of the matrix treated as a contingency table.
A value of $\phi = 0$ means the correlation is random and $\phi = 1$ is a perfect fit (incidentally, $\phi = -1$ mean complete anti-correlation); thus we can use $\phi$ as a measure of confidence (avoiding false positives and false negatives) in addition to the na\"{\i}ve precision.

\end{description}

\subsection{Initial Experiments}
Armed with the appropriate mathematical data and the technique of ML, implemented via either Python's Keras/TensorFlow \cite{python} or Wolfram's Mathematica version $>11.0$\cite{mathematica}.
One can take a simple MLP of the form (the hyper-parameters and architecture differ for specific cases and the following is only an illustration of a typical case)
\begin{equation}\label{NN}
\begin{array}{c}\includegraphics[trim=10mm 0mm 0mm 0mm, clip, width=5in]{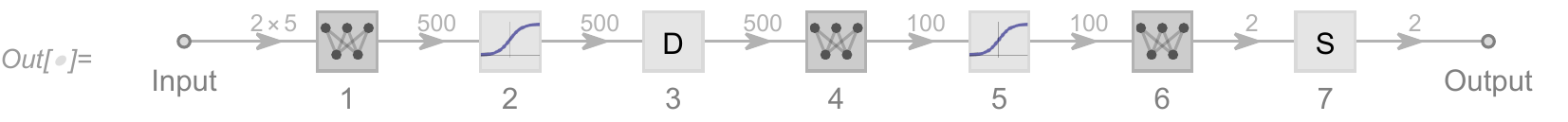}\end{array}
\end{equation}
where the hidden layers are
(1) a fully-connected linear layer of 500 nodes; (2) element-wise sigmoid activation $\sigma(z) := (1 + e^{-z})^{-1}$;
(3) dropout layer (we switch off neurons with some probability in order not to over-fit); (3) linear layer of 100  nodes; (4) sigmoid; (5) linear layer of 2 nodes; (6) Softmax to output.
Such an NN was found to estimate the size of Hodge numbers for CICYs and WP4 hypersurfaces very well in a matter of seconds on an ordinary laptop \cite{He:2017aed}.
One could think of this setup, a fully connected neural networks of depth $d$,
as the following composition of maps:
\begin{equation}
 \mathbb{R}^{n_0}\stackrel{L_{n_1}}{\longrightarrow}\mathbb{R}^{n_1}\stackrel{f}{\longrightarrow}\mathbb{R}^{n_1}\stackrel{L_{n_2}}{\longrightarrow}\cdots\stackrel{L_{n_d}}{\longrightarrow}\mathbb{R}^{n_d} \rightarrow \IR 
 \; ,
\end{equation} 
where $L_n$ are activation functions (such as sigmoids) with trainable weights and biases and co-domain dimension $n$ and the last layer outputs some real value or discrete value.
The power of the MLP is the harnessing of the ultimately complicated (not even necessarily analytic) structure of the composite map.

A detailed analysis was carried out in \cite{Bull:2018uow} where the 19-way classifier/regressor in an architecture similar to \eqref{NN} as well as an SVM, performed to about 90\% accuracy in an 80-20\% training-validation split \footnote{
We remark that for CICY 4-folds, where the data is about a million, accuracies to around 96\% was achieved
\cite{He:2020lbz}.
}.
It is interesting that in these experiments, one never exploited the {\it matrix} structure of the input: e.g., the CICY configuration was flattened a long vector of integers.
This is quite contrary to the image processing of \eqref{sampledigits} where a convolutional network (CNN) would be used which ``convolves'' with nearest neighbours.
Such CNNs were indeed tried more recently and $>99\%$ accuracies were reached \cite{Erbin:2020tks}.

\subsection{More Success Stories in String/Geometry}
 One can imagine that all computational problems in string phenomenology and more generally in computational algebraic geometry could benefit from the paradigm of machine-learning.
 Indeed, the initial explorations of \cite{Krefl:2017yox} on Calabi-Yau volumes, of \cite{Ruehle:2017mzq} on line-bundle cohomology, and of \cite{Carifio:2017bov} on F-theory compactifications, in conjunctions with \cite{He:2017aed}, launched the String Data conference series from 2017 on-wards.
 
Though it is difficult to review all the works since, I will give a bird's-eye-view of the the various directions taken, first within string/geometry, and then more generally to other branches of mathematics.
In the former, some major directions and success stories (with $0.90$ accuracies with relatively simple architectures) have included

\begin{description}

\item[Heterotic: ] Selection of MSSM from heterotic orbifold constructions \cite{Parr:2019bta,Parr:2020oar}, distinguishing standard models from heterotic line bundles \cite{Otsuka:2020nsk,Deen:2020dlf}.
Machine-learning of bundle cohomology of surfaces \cite{Brodie:2019dfx} as well as toric hypersurfaces \cite{Klaewer:2018sfl}. One points out \cite{Constantin:2018hvl} where exact formulae were found for line-bundle cohomology through an MLP exploration of the regions in moduli space.

\item[F/M-Theory: ]  Finding gauge groups \cite{Wang:2018rkk} and matter-content \cite{Bies:2020gvf} within F-theory compactifications. Distinguishing elliptically fibered manifolds within the CICYs \cite{He:2019vsj}.
Decidability issue of diophantine systems in K\"ahler stabilization \cite{Halverson:2019vmd}.

\item[Type II: ] topological data analysis \cite{Cole:2018emh} of, and genetic algorithms for searching within \cite{Cole:2019enn} flux vacua in type II.
Reinforcement learning explorations of IIA brane configurations \cite{Halverson:2019tkf} and IIB landscape \cite{1830622}.
Seiberg duality in type IIB quiver theories \cite{Bao:2020nbi}.

\item[Physical Symmetries: ] symmetries in various physical systems (including representations of CICYs) \cite{Krippendorf:2020gny}, and CFT symmetries \cite{Chen:2020dxg}.

\item[Metric: ] As mentioned several times, there is no known  analytic Calabi-Yau metric on a non-trivial compact K\"ahler manifold. Donaldson developed an efficient numerical algorithm using the method of balanced metrics from a potential formed by increasing powers of monomial sections \cite{metricDon}, which were then nicely implemented in \cite{Douglas:2006rr,Braun:2007sn} (q.v..~also the functional method of \cite{Headrick:2005ch}).
It was shown in \cite{Ashmore:2019wzb} that Donaldson's algorithm can be machine-learnt (and to 10-100-fold increase in efficiency).

\item[Cosmology: ] The cosmic landscape \cite{Liu:2017dzi}, especially vacuum selection from cosmology constraints \cite{Carifio:2017nyb}, were studied. Interesting network structures were found in \cite{Carifio:2017nyb} and
\cite{Khoury:2019ajl} studied certain accessibility measures in inflation and \cite{Rudelius:2018yqi}, machine learning in inflation.

\item[``Meta'' Physics: ] A fun experiment was undertaken in \cite{He:2018dlv} where all titles from hep-th and four related sections of the arXiv: hep-ph, hep-lat, gr-qc, and math-ph were downloaded since the beginning and fed into the NN {\it Word2Vec} (about $10^6$ titles). 
Interesting linear syntactical identities such as ``holography + quantum + string + ads = extremal-black-hole'' presented themselves and the syntactical structure of the different sections were indeed found to be distinct.
 
\end{description}

Of particular note are the striking ideas in \cite{Hashimoto:2018ftp,Hashimoto:2019bih,Koch:2019fxy,Halverson:2020trp} where the fundamentals of quantum field theory, holography and renormalization group flow, are phrased in terms of appropriate neural networks.
Indeed, the reader is also referred to the recent works of \cite{vanchurin,cellular} on the possible computational nature of reality itself.

\section{Outlook: ML Mathematical Structures}
Given the efficacy of ML in so many directions in string/geometry, it is natural to ask whether and how different problems in mathematics respond to ML.
We leave a detailed discussion of this to \cite{MLmaths}, but for now, it is perhaps fitting that we conclude this talk with some conducive experiments which have been performed as well as some speculations for the future.
Let us approximately group the successful experiments by subject:

\begin{description}
\item[Algebraic Geometry over $\IC$: ]
Most of the problems mentioned above fall under this heading.
We need to emphasize that we work over $\IC$, an algebraically closed number field. 
Any problem in computational algebraic geometry essentially boils down to finding kernels and co-kernels of integer matrices (in appropriate monomial bases), something quite adaptable to ML.
A recent work on using reinforcement learning to perform the key step of finding S-pairs in constructing Gr\"obner bases was done in \cite{mikeGB}.

\item[Representation Theory: ]
Preliminary investigation on whether SVMs and MLPs can distinguish finite groups and finite rings from random matrix structures was initiated in \cite{He:2019nzx}; more surprising was the fact that {\it simple groups} seemed to be distinguishable.
For continuous groups, lengths of branching rules and tensor decomposition in Lie algebras can also be learned  by looking at weight vectors \cite{Chen:2020jjw}; this is obviously also of importance to particle physics.

\item[Knot Theory: ]
Jones polynomials and complementary volume of knots are studied from ML in \cite{Jejjala:2019kio} and letting ML find configurations of knots themselves, in \cite{Gukov:2020qaj}.

\item[Graph Theory and Combinatorics: ]
Cluster mutation on quivers was studied in \cite{Bao:2020nbi}.
On a more basic level, properties of finite simple graphs, such as whether it possesses Euler or Hamilton cycles, whether it is flat (there is a notion of Ricci-flatness for finite graphs), etc. were studied in \cite{He:2020fdg}.

\item[Number Theory: ]
As one might imagine, a direct attack on predicting the next prime number by ML is most likely unfruitful \cite{He:2017aed,He:2018jtw}. Likewise, predicting quantities relevant to the Birch-Swinnerton-Dyer Conjecture was also difficult \cite{Alessandretti:2019jbs}.
Surprisingly, however, problems in arithmetic geometry, ranging from dessins d'enfant \cite{He:2020eva} ($>0.9$ accuracy), to arithmetic properties of hyper-elliptic curves \cite{He:2020kzg} ($\sim 0.99 - 1.00$ accuracies) and Galois number field extensions of the rationals \cite{MLfields} ($>0.9$ accuracy) behaved very well to simple classifiers such as Na\"{\i}ve Bayes.

\item[Symbolic Manipulation: ]
Recent advances in generating new identities in calculus \cite{DLsymb} and continued fractions \cite{ramanujanML} have met with impressive success. So too, have there been tools to extract fundamental laws  \cite{scinet} and formulae \cite{Udrescu:2019mnk} of physics.
\end{description}

With these tantalizing thoughts let us conclude my talk here.
We have seen how into the alembic of mathematics and fundamental physics is now infused, over the last few years, new techniques of the data revolution, especially the predictive power of machine-learning and neural networks.
We are, of course, only at the early stage. Having an ML predict a result, even to 100\% accuracy, does not always mean one could obtain analytic information as to why. What we hope for, is what the physicist Max Tegmark calls ``intelligible intelligence'', where we can formulate new results, or at least conjecture precise statements, when we are given an ML algorithm which performs superbly well. When I shared my initial excitement back in 2017, of the prospects to machine-learn problems ranging from geometry to algebra, to my friend the logician Boris Zilber, he astutely remarked:
``now you have syntax, it would be good to find the semantics.''

\section*{Acknowledgments}
We are grateful for the kind invitations, in person and over Zoom, of the various institutions over this most extraordinary year of 2020 -- the hospitality and conversations before the lock-down and the opportunity for a glimpse of the outside world during: 
Harvard University, Tsinghua University/BIMSA, Universidad Cat\'{o}lica del Norte Chile, London Institute of Mathematical Sciences, Queen's Belfast, King's College London, University of Connecticut, ``Clifford Algebra \& Applications 2020'' at UST China, ``String Maths 2020'' at Capetown, ``International Congress Mathematical Software 2020'' at Braunschweig,  University of Torino, ``SageMath/M2 - an Open Source Initiative'' at the University of Minnesota, ``East Asia Strings'' at Taipei-Seoul-Tokyo, Nankai University, Imperial College London, and Nottingham University.
The work, as always, is indebted to STFC UK for grant ST/J00037X/1 and Merton College, Oxford for a quiet corner of paradise.


\end{document}